\documentclass[12pt]{elsart}
\usepackage{latexsym}
\usepackage{epsfig}
\usepackage{amsmath}

\newcommand{\be}{\begin{equation}}
\newcommand{\ee}{\end{equation}}
\newcommand{\br}{\begin{eqnarray}}
\newcommand{\er}{\end{eqnarray}}

\newcommand{\TM}{T_{\mathrm{M}}}
\newcommand{\fm}{f_{\mathrm{m}}}
\newcommand{\zetam}{\zeta_{\mathrm{m}}}
\newcommand{\aM}{a_{\mathrm{M}}}
\newcommand{\Zm}{Z_{\mathrm{m}}}
\newcommand{\pe}{p_{\mathrm{e}}}

\newcommand{\mm}{m_{\mathrm{m}}}
\newcommand{\cm}{c_{\mathrm{m}}}
\newcommand{\km}{k_{\mathrm{m}}}
\newcommand{\Mp}{M_{\mathrm{p}}}
\newcommand{\Cp}{C_{\mathrm{p}}}
\newcommand{\Kp}{K_{\mathrm{p}}}
\newcommand{\Ca}{C_{\mathrm{a}}}
\newcommand{\Ka}{K_{\mathrm{a}}}
\newcommand{\Lp}{L_{\mathrm{p}}}
\newcommand{\La}{L_{\mathrm{a}}^{n}}
\newcommand{\Mf}{M_{\mathrm{f}}}
\newcommand{\Mfs}{M_{\mathrm{f}}^{\mathrm{s}}}
\newcommand{\Ds}{D_{\mathrm{s}}}

\newcommand{\mass}{\ \mathrm{g}\cdot\mathrm{cm}^{-2}}
\newcommand{\stif}{\ \mathrm{dyn}\cdot\mathrm{cm}^{-3}}
\newcommand{\damp}{\ \mathrm{dyn}\cdot\mathrm{s}\cdot\mathrm{cm}^{-3}}

\renewcommand{\theequation}{\arabic{section}.\arabic{equation}}

\begin{document}

\begin{frontmatter}

\title{Signal processing of acoustic signals in the time domain with an active nonlinear nonlocal 
cochlear model}

\author{M. Drew LaMar\thanksref{MDLemail}}
\address{Department of Mathematics, University of Texas at Austin,
Austin, TX 78712, USA.}
\thanks[MDLemail]{Corresponding author (mlamar@math.utexas.edu)}
\author{Jack Xin}
\address{Department of Mathematics and ICES (Institute of Computational
Engineering and Sciences), University of Texas at Austin, Austin, TX 78712, USA.}
\author{Yingyong Qi}
\address{Qualcomm Inc,
5775 Morehouse Drive,
San Diego, CA 92121, USA.}

\date{}
\setcounter{page}{1}
\setcounter{section}{0}
\begin{abstract}
A two space dimensional active nonlinear nonlocal cochlear model is 
formulated in the time domain 
to capture nonlinear hearing effects such as compression, 
multi-tone suppression and difference tones.  The 
micromechanics of the basilar membrane (BM) are incorporated to model 
active cochlear properties.  An active gain parameter
is constructed in the form of 
a nonlinear nonlocal functional of BM displacement.  
The model is discretized with 
a boundary integral method and  
numerically solved using an iterative second order accurate 
finite difference scheme.  
A block matrix structure of the discrete 
system is exploited to simplify the numerics with no loss
of accuracy.  Model responses to multiple frequency stimuli 
are shown in agreement with hearing experiments. A nonlinear spectrum 
is computed from the model, and compared with FFT spectrum for noisy 
tonal inputs.
The discretized model
is efficient and accurate, and can serve as a useful
auditory signal processing tool.
\end{abstract}


\end{frontmatter}

\newpage

\section{Introduction}
\setcounter{equation}{0}
Auditory signal processing based on phenomenological models of 
human perception has helped to 
advance the modern technology of 
audio compression \cite{pohlmann}.  
It is of interest therefore to 
develop a systematic mathematical framework 
for sound signal processing based on models of the ear.
The biomechanics of 
the inner ear (cochlea) 
lend itself well to mathematical 
formulation (\cite{allen,deboer} among others). Such  
models can recover main aspects of the physiological data  
\cite{bekesy,ruggero} for simple acoustic inputs (e.g. single frequency tones). 
In this paper, we study a nonlinear nonlocal model and associated numerical 
method for processing complex signals (clicks and noise) in the time domain. 
We also obtain a new spectrum of sound signals with nonlinear hearing 
characteristics which can be of potential interest for applications such 
as speech recognition.   

Linear frequency domain cochlear models have been around for a long time and studied 
extensively \cite{neely,neelykim}.  The cochlea, however, is known to have 
nonlinear characteristics, such as compression, two-tone suppression and combination tones, which
are all essential to capture interactions of multi-tone complexes \cite{dengg,dengk,jxintime}.  In this
nonlinear regime, it is more expedient to work in the time domain to resolve complex nonlinear frequency
responses with sufficient accuracy.  The nonlinearity in our model
resides in the outer hair cells (OHC's), which act as an amplifier to 
boost basilar membrane (BM) responses to low-level
stimuli, so called active gain.  It has been shown \cite{deboernut} that this type of nonlinearity is also nonlocal in
nature, encouraging near neighbors on the BM to interact.

One space dimensional transmission line models with nonlocal nonlinearities have been studied previously
for auditory signal processing \cite{dengk,jxinglobal,jxintime,jxinPDE}. 
Higher dimensional models 
give sharper tuning curves and higher frequency selectivity.
In section 2, we begin with a two space dimensional (2-D) 
macromechanical partial differential equation (PDE) model. 
We couple the 2-D model with the BM micromechanics
of the active linear system in \cite{neelykim}. 
We then make the gain parameter nonlinear
and nonlocal to complete the model setup, and 
do analysis to simplify the model.

In section 3, we discretize the system and formulate a second order accurate
finite difference scheme so as to combine efficiency and accuracy.  The matrix we need to invert at 
each time step has a time-independent part (passive) and a time-dependent part (active).  In order
to speed up computations, we split the matrix into the passive 
and active parts and devise an
iterative scheme.  We only need to invert the passive part once, thereby significantly speeding
up computations.  The structure of the system also allows us to reduce the complexity of the problem
by a factor of two, giving even more computational efficiency.  A proof of convergence of the iterative scheme is 
given in the Appendix.

In section 4, 
we discuss numerical results and show that 
our model successfully reproduces the nonlinear effects such as compression,
 multi-tone suppression, and combination difference tones.  We demonstrate such effects by inputing
various signals into the model, 
such as pure tones, clicks and noise. A nonlinear spectrum is 
computed from the model and compared with FFT spectrum for the acoustic input of a single tone 
plus Gaussian white noise. The conclusions are in section 5.

\begin{figure}[tb]
\epsfig{file=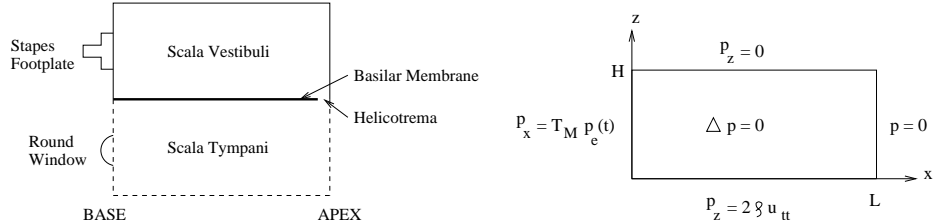,width=350pt}
\caption{The figure on the left is a schematic of the cochlea, while the figure on the right 
represents the upper chamber with the macromechanical equations and boundary conditions.}
\label{fig:coch}
\end{figure}

\section{Model Setup}

\subsection{Macromechanics}

The cochlea consists of an upper and lower fluid filled chamber, the scala vestibuli and scala tympani, 
with a shared elastic boundary called the basilar membrane (BM) (see Figure \ref{fig:coch}).  The BM acts like a Fourier 
transform with each location on the BM tuned to resonate at a particular frequency, ranging 
from high frequency at the basal end to low frequency at the apical end.  The acoustic wave 
enters the ear canal, where it vibrates the eardrum and then is filtered through the middle ear, 
transducing the wave from air to fluid in the cochlea via the stapes footplate.  A traveling
wave of fluid moves from the base to the apex, creating a skew-symmetric motion of the BM.  
The pressure difference drives the BM, which resonates according to the frequency content of the passing wave.

We start with simplification of the upper cochlear chamber into a two dimensional rectangle
$\Omega = [0,L] \times [0,H]$ (see Figure \ref{fig:coch}). Due to the symmetry, we can ignore the lower
chamber.  The bottom boundary ($z = 0$) is the BM, while the left boundary ($x = 0$) is the stapes 
footplate.  The macromechanical equations are
\begin{equation}
  \left\{ \begin{array}{ll}
             \varDelta p(x,z,t) = \frac{\partial^2 p}{\partial x^2} + \frac{\partial^2 p}{\partial z^2} = 0,\ x \in [0,L],\ z \in [0,H],\ t \in [0,\infty) \\
	     p_{x}(0,z,t) = \TM\pe(t),\ \ p(L,z,t) = 0 \\
             p_{z}(x,0,t) = 2\rho u_{tt},\ \ p_{z}(x,H,t) = 0
           \end{array}
  \right.
  \label{eq:macro}
\end{equation}
where $p(x,z,t)$ is the pressure difference across the BM, $u(x,t)$ denotes BM displacement, and 
$\rho$ is fluid density.

At the stapes footplate $(0,z)$, $\pe(t)$ is pressure at the eardrum while $\TM$ is a bounded 
linear operator on the space of bounded continuous functions that incorporates the middle ear 
filtering characteristics.  In the frequency domain, for each input $\e^{\mathrm{i}\omega t}$, 
$\TM(\omega) = 2\rho \mathrm{i}\omega/\Zm$, where $\Zm$ is the impedance of the middle ear.  
The middle ear amplification function is given by $\aM = |\TM|$.  In our case, based on Guinan and Peake 
\cite{guinan},
\begin{equation}
\aM(f) = 1.815f^2((1-\frac{f^2}{\fm^{2}})^2 + (2\zetam f/\fm)^2)^{-1/2},
\label{eq:ssmear}
\end{equation}
where $\fm = 4\ \mathrm{kHz}$ is the middle ear characteristic frequency and $\zetam = 0.7$ is the middle
ear damping ratio.  Thus, for $\pe(t) = A \exp\{2\pi \mathrm{i}f\} + c.c$, where c.c. is complex conjugate, we
have $\TM\pe(t) = B \exp\{2\pi \mathrm{i}f\} + c.c.$, where $B = \aM(f)A$.

For more complex stimuli, it is useful to model the middle ear in the time domain as a one-degree of freedom
spring-mass system.  The equivalent time domain formulation of the steady state middle ear is given by
\begin{equation}
  \left\{ \begin{array}{l}
             \pe(t) = \mm\ddot{s}(t) + \cm\dot{s}(t) + \km s(t) \\
             s(0) = \dot{s}(0) = 0
	  \end{array}
  \right.
  \label{eq:tdmear}
\end{equation}
where $s(t)$ is stapes displacement and $\mm$, $\cm$ and $\km$ are the mass, damping and stiffness of the middle ear.
The stapes boundary condition in (\ref{eq:macro}) is replaced by
\begin{equation}
  p_{x}(0,z,t) = 2\rho\ddot{s}(t)
  \label{eq:bcmear}
\end{equation}
One of the interesting effects of using the time domain middle ear model 
is that it reduces the dispersive instability in the 
cochlea (see \cite{jxindisp}).  It appears that 
the steady state middle ear model ignores important transient effects and
phase shifts that help to reduce the shock to the cochlea.

At the helicotrema $(L,z)$, we have used the 
Dirichlet boundary condition $p(L,z,t) = 0$.  In \cite{neelykim},
they used an absorbing boundary condition $p_{x}(L) = cp_{t}(L)$, where $c$ is a positive constant \cite{neelykim}.
Other models use the Neumann condition $p_{x}(L) = 0$.  It has been stated that the frequency domain solutions
are minimally affected by which boundary condition is chosen \cite{neely}, and thus we have chosen the simpler Dirichlet condition.  
However, interesting results on choosing the best initial conditions to minimize transient effects (dispersive
instability) has been shown in \cite{jxindisp} using the Neumann condition.  
To summarize, the macromechanics consist of equations (\ref{eq:macro})--(\ref{eq:bcmear}). 
\bigskip

\subsection{Micromechanics}

\begin{figure}[tb]
\parbox{2in}{\epsfig{file=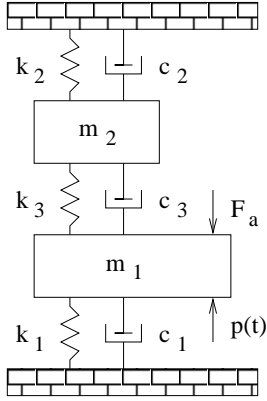,width=100pt}}
\parbox{3in}{\caption{Cross section micromechanics of the cochlea.  The mass $m_{1}$ represents a cross section
of the BM, while mass $m_{2}$ is a cross section of the TM.  (Reconstructed from Figure 3 in \cite{neelykim})}\label{fig:sm}}
\end{figure}

This is a resonant tectorial membrane model based on  
\cite{neelykim}.  The BM and TM 
(tectorial membrane) are modeled 
as two lumped masses coupled by a spring and damper, with each mass connected to a wall
by a spring and damper.  (See Figure \ref{fig:sm}).  A classical approximation
is to have no longitudinal coupling except that which occurs through the fluid.  Denoting $\xi(x,t) = (u(x,t),v(x,t))$
as BM and TM displacement, respectively, the equations of motion for the passive case at each point along the 
cochlea are given by
\begin{equation}
\Mp\ddot{\xi} + \Cp\dot{\xi} + \Kp\xi = F
\label{eq:passive}
\end{equation}
where
\begin{equation}
  \Mp = \left[\begin{array}{cc}
               m_{1} & 0 \\
	       0     & m_{2}
	    \end{array}
      \right], \ \
  \Cp = \left[\begin{array}{cc}
               c_{1}+c_{3} & -c_{3} \\
	       -c_{3} & c_{2}+c_{3}
	    \end{array}
      \right], \ \
  \Kp = \left[\begin{array}{cc}
               k_{1}+k_{3}  & -k_{3} \\
	       -k_{3} & k_{2}+k_{3}
	    \end{array}
      \right]
\label{eq:passmat}
\end{equation}
and forcing function
\begin{equation}
      F = \left[\begin{array}{c}
                   p(x,0,t) \\
                   0
                \end{array}
          \right]
\label{eq:force}
\end{equation}
The parameters $m_{i}$, $c_{i}$, and $k_{i}$ are functions of $x$.  The initial conditions are given by
\begin{equation}
  \xi(x,0) = \dot{\xi}(x,0) = 0
  \label{eq:ic}
\end{equation}
To make the model active, a self-excited vibrational force acting on the BM is added to (\ref{eq:passive}):
\[
\Mp\ddot{\xi} + \Cp\dot{\xi} + \Kp\xi = F + F_{\mathrm{a}}
\]
where
\[
      F_{\mathrm{a}} = \left[\begin{array}{c}
                   \gamma[c_{4}(\dot{u} - \dot{v}) + k_{4}(u-v)] \\
                   0
                \end{array}
          \right]
\]
The difference $u-v$ represents OHC displacement.
The parameter $\gamma \in [0,1]$ is the active gain control.  In \cite{neelykim}, this is a constant, but in our 
case will be a nonlinear nonlocal functional of BM displacement and BM location.  Bringing $F_{\mathrm{a}}$ to the left, we have
\begin{equation}
  \Mp\ddot{\xi} + (\Cp-\gamma \Ca)\dot{\xi} + (\Kp-\gamma \Ka)\xi = F
  \label{eq:active}
\end{equation}
where
\begin{equation}
  \Ca = \left[\begin{array}{cc}
               c_{4} & -c_{4} \\
	       0     & 0
            \end{array}
      \right], \ \
  \Ka = \left[\begin{array}{cc}
               k_{4} & -k_{4} \\
	       0     & 0
	    \end{array}
      \right]
  \label{eq:actmat}
\end{equation}
Thus, the micromechanics consist of equations (\ref{eq:passmat})--(\ref{eq:actmat}).

\subsection{Nonlinear Nonlocal Active Gain}
A compressive nonlinearity in the model is necessary to capture effects such as two-tone suppression and combination
tones.  Also, to allow for smoother BM profiles, we make the active gain nonlocal.  Thus we have
\[
  \hat{u}(x,t) = \frac{2}{\sqrt{\lambda\pi}}\int_{0}^{L}\e^{-(x-s)^2/\lambda}u^{2}(s,t)\ \d s
\]
and gain
\[  
  \gamma(x,t) = \frac{1}{1+\theta\hat{u}}
\]
where $\theta,\ \lambda$ are constants.

\subsection{Semi-discrete Formulation}
Solving the pressure Laplace equation on the rectangle using separation of variables, we arrive at
\begin{equation}
  p(x,0,t) = \TM\pe(t)(x-L) + \sum_{n=1}^{\infty}A_{n}\cos\beta_{n}x
  \label{eq:sov}
\end{equation}
where
\begin{equation}
  A_{n} = \left(\frac{-4\rho H}{L}\right)\left(\frac{\coth\beta_{n}H}{\beta_{n}H}\right)\int_{0}^{L}u_{tt}(x,t)\cos\beta_{n}x\d x
  \label{eq:integral}
\end{equation}
\[\beta_{n} = \frac{(n-\frac{1}{2})\pi}{L}\]
Substituting (\ref{eq:sov}) into (\ref{eq:force}) and then discretizing (\ref{eq:active}) in space into $N$ grid points, we have
\begin{equation}
  M\vec{\xi}_{tt} + C(t)\vec{\xi}_{t} + K(t)\vec{\xi} = \vec{b}(t)
  \label{eq:disc}
\end{equation}
where
\[M = \left[\begin{array}{cc}
               M_{1}+\alpha \Mf & 0 \\
               0                  & M_{2}
            \end{array}
      \right]
\]
\[C(t) = \Cp - \hat{\Gamma}(t) \Ca = \left[\begin{array}{cc}
                                      C_{1}+C_{3}-\Gamma(t) C_{4} & -(C_{3}-\Gamma(t) C_{4}) \\
                                      -C_{3} & C_{2}+C_{3}
                                   \end{array}
                             \right]
\]
\[K(t) = \Kp - \hat{\Gamma}(t) \Ka = \left[\begin{array}{cc}
                                      K_{1}+K_{3}-\Gamma(t) K_{4} & -(K_{3}-\Gamma(t) K_{4}) \\
                                      -K_{3} & K_{2}+K_{3}
                                   \end{array}
                             \right]
\]
\[\vec{b}(t) = \left[\begin{array}{c}
                     \TM\pe(t)(\vec{x}-L) \\
		     0
		  \end{array}
            \right]
\]
\[M_{\mathrm{f},ij} = \sum_{k=1}^{K}\frac{\coth\beta_{k}H}{\beta_{k}H}\cos(\beta_{k}x_{i})\cos(\beta_{k}x_{j})w_{j}\]
\[\alpha = \frac{4\rho H}{N-1}\]
$\Cp$, $\Kp$, $\Ca$ and $\Ka$ are now block diagonal, where $K_{i} = \hbox{diag}\{k_{i}\}$ and $C_{i} = \hbox{diag}\{c_{i}\}$.
Also, $M_{i} = \hbox{diag}\{m_{i}\}$, $\Gamma(t) = \hbox{diag}\{\gamma_{i}(t)\}$ and $\hat{\Gamma}(t) = \hbox{diag}\{\Gamma(t),0\}$.
The numbers $w_{j}$ are numerical integration weights in the discretization of (\ref{eq:integral}) and are chosen based on the 
desired degree of accuracy.
Note that we can write $\Mf = \Mfs W$, where $W = \hbox{diag}(w_{j})$
and $\Mfs$ is symmetric and positive definite.  The result of separation of variables produced the matrix $\Mf$, 
which is essentially the mass of fluid on the BM and {\em dynamically couples} the system.

\section{Numerics}
In formulating a numerical method, we note that the matrices in (\ref{eq:disc}) can be split into a time-independent passive
part and a time-dependent active part.  In splitting in this way, we are able to formulate an iterative scheme where we only
need to do one matrix inversion on the passive part for the 
entire simulation.  Thus, using second order approximations of the 
first and second derivates in (\ref{eq:disc}), we arrive at
\begin{equation}
  (\Lp-\La)\vec{\xi}^{n+1} = \vec{B}^{n} \;\;\; \Longrightarrow \;\;\; \vec{\xi}^{n+1,k+1} = \Lp^{-1}\vec{B}^{n} + \Lp^{-1}\La\vec{\xi}^{n+1,k}
  \label{eq:iteration}
\end{equation}
where superscript $n$ denotes discrete time, $k$ denotes iteration and
\[\Lp = 2M + \frac{3}{2}\Delta{t}\Cp + \Delta{t}^{2}\Kp\]
\[\La = \hat{\Gamma}^{n}[\frac{3}{2}\Delta{t}\Ca + \Delta{t}^{2}\Ka]\]
\[\vec{B}^{n} = \Delta{t}^{2}\vec{b}(n\Delta{t}) + 
    M(5\vec{\xi}^{n} - 4\vec{\xi}^{n-1} + \vec{\xi}^{n-2}) + 
    \frac{\Delta{t}}{2}C^{n}(4\vec{\xi}^{n} - \vec{\xi}^{n-1}).\]
Proof of convergence will follow naturally from the next discussion.  Notice that this is a 
$2N \times 2N$ system.  We shall simplify it to an $N \times N$ system 
and increase the computational efficiency.  

\subsection{System Reduction}

We write $\Lp$ and $\La$ in block matrix form as
\[\Lp = \left[\begin{array}{cc}
                   \tilde{M}_{1} & -P_{3} \\
		   -P_{3}        & \tilde{M}_{2}
		\end{array}
	  \right]
\]
\[\La = \left[\begin{array}{cc}
                   \Gamma^{n} P_{4} & -\Gamma^{n} P_{4} \\
		   0            & 0
		\end{array}
          \right]
\]
where
\[\tilde{M}_{1} = 2(\alpha \Mf + M_{1}) + P_{1} + P_{3}\]
\[\tilde{M}_{2} = 2M_{2} + P_{2} + P_{3}\]
\[P_{i} = \frac{3}{2}\Delta{t}C_{i} + \Delta{t}^{2}K_{i}\]
It is easily seen that the left inverse of $\Lp$ is given by
\[
    \Lp^{-1} = \left[\begin{array}{cc}
                        D^{-1} & D^{-1}\tilde{M}_{2}^{-1}P_{3} \\
			\tilde{M}_{2}^{-1}P_{3}D^{-1} & \tilde{M}_{2}^{-1}P_{3}D^{-1}\tilde{M}_{1}P_{3}^{-1}
	             \end{array}
	       \right]
\]
where 
\begin{eqnarray}
  D & = & \tilde{M}_{1} - P_{3}\tilde{M}_{2}^{-1}P_{3} \nonumber \\
    & = & 2\alpha \Mf + [2M_{1} + P_{1} + P_{3}(I - \tilde{M}_{2}^{-1}P_{3})] \nonumber \\
    & = & \{2\alpha \Mfs + [2M_{1} + P_{1} + P_{3}(I - \tilde{M}_{2}^{-1}P_{3})]W^{-1}\}W \nonumber \\
    & \equiv & \Ds W
  \label{eq:dsdef}
\end{eqnarray}
Note that $D$ is invertible since $\Mfs$ is positive definite, thus invertible, and all other terms are positive diagonal
matrices, and thus their sum is positive definite and invertible.  We also have
\begin{equation}
      \Lp^{-1}\La = \left[\begin{array}{cc}
                             D^{-1}\Gamma^{n} P_{4} & -D^{-1}\Gamma^{n} P_{4} \\
			     \tilde{M}_{2}^{-1}P_{3}D^{-1}\Gamma^{n} P_{4} & -\tilde{M}_{2}^{-1}P_{3}D^{-1}\Gamma^{n} P_{4}
			  \end{array}
	            \right]
      \label{eq:lpinvla}
\end{equation}
Letting $\vec{B}^{n} = (\vec{B}_{1}^{n},\vec{B}_{2}^{n})$, we have
\begin{eqnarray}
  \vec{u}^{n+1,k+1} & = & W^{-1}[\zeta_{1}^{n} + \Ds^{-1}\Gamma^{n} P_{4}(\vec{u}-\vec{v})^{n+1,k}] \label{eq:iter1} \\
  \vec{v}^{n+1,k+1} & = & W^{-1}\tilde{M}_{2}^{-1}P_{3}[\zeta_{2}^{n} + \Ds^{-1}\Gamma^{n} P_{4}(\vec{u}-\vec{v})^{n+1,k}] \label{eq:iter2}
\end{eqnarray}
where
\begin{equation}
  \zeta_{1}^{n} = \Ds^{-1}[\vec{B}_{1}^{n} + \tilde{M}_{2}^{-1}P_{3}\vec{B}_{2}^{n}] \label{eq:zeta1}
\end{equation}
\begin{equation}
  \zeta_{2}^{n} = \Ds^{-1}[\vec{B}_{1}^{n} + \tilde{M}_{1}P_{3}^{-1}\vec{B}_{2}^{n}] \label{eq:zeta2}
\end{equation}
At each time step, we do 2 $N \times N$ matrix solves in (\ref{eq:zeta1}) and (\ref{eq:zeta2}) to initialize the 
iterative scheme.  Then, since the same term appears in both equations (\ref{eq:iter1}) and (\ref{eq:iter2}), for 
each $k$ we only have to do 1 $N \times N$ matrix solve.  In practice, since $\Ds$ is symmetric,
positive definite and {\em time-independent}, we compute the Cholesky factorization of $\Ds$ at the start of the simulation
and use the factorization for more efficient matrix solves at each step.  As a side note, if we subtract (\ref{eq:iter2}) from 
(\ref{eq:iter1}), we have one equation for the OHC displacement $u-v$.

\section{Numerical Results}

\subsection{Model Parameters}
We start with a modification of the parameters in \cite{neelykim} (See Table \ref{tab:params}).
It is known that higher dimensional models give higher sensitivity.  This is the case with this model.  The 1-D
 model \cite{neelykim} gives a 90 dB active gain at 16 kHz, whereas the 2-D model gives a 160 dB active gain.  Thus, we need to
tune the system to reduce the gain.  There are many ways to do this, and the method we choose is to increase all the damping
coefficients in the table by the following:
\[2\e^{0.2773x}c_{i} \mapsto c_{i},\ \ i=1,2,3,4\]

\begin{table}[tb]
\footnotesize
\caption{Model parameters in cgs units}
\label{tab:params}
\begin{center}
\begin{tabular}{||l|l||l|l||}	\hline
$m_{1}(x)$	&	$3\cdot10^{-3}\mass$		&	$\mm$		&	$34.4\cdot10^{-3}\mass$	\\
$c_{1}(x)$	&	$20+1500\e^{-2x}\damp$		&	$\cm$		&	$1.21\cdot10^{3}\damp$	\\
$k_{1}(x)$	&	$1.1\cdot10^{9}\e^{-4x}\stif$	&	$\km$		&	$2.18\cdot10^{7}\stif$	\\
$m_{2}(x)$	&	$0.5\cdot10^{-3}\mass$ 		&	$L$		&	$2.5$ cm		\\
$c_{2}(x)$	&	$10\e^{-2.2x}\damp$		&	$H$		&	$0.1$ cm		\\
$k_{2}(x)$	&	$7\cdot10^{6}\e^{-4.4x}\stif$	&	$\rho$		&	$0.1 \ \mathrm{g}\cdot\mathrm{cm}^{-3}$		\\
$c_{3}(x)$	&	$2\e^{-0.8x}\damp$ 		&	$\theta$	&	$0.5$		\\
$k_{3}(x)$	&	$10^{7}\e^{-4x}\stif$ 		&	$\lambda$	&	$0.08$ cm		\\
$c_{4}(x)$	&	$1040\e^{-2x}\damp$ 		&	$\Delta t$	&	$2.5\cdot10^{-6}$ -- $10^{-5}$ s\\
$k_{4}(x)$	&	$6.15\cdot10^{8}\e^{-4x}\stif$	&	$N$		&	$401$		\\ \hline
\end{tabular}
\end{center}
\end{table}

\subsection{Isointensity Curves}

\begin{figure}[tb]
\centerline{\epsfig{file=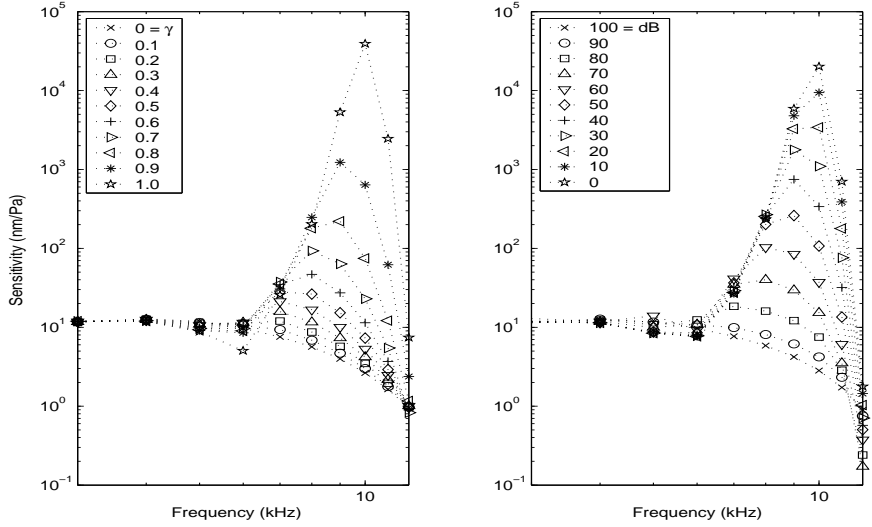,width=330pt,height=200pt}}
\caption{Both figures are sensitivity curves for CP = 0.77 cm or CF = 10 kHz.  The left plot is a collection of
sensitivity curves for the linear steady state active model where the parameter is the active gain $\gamma$.  The right
plot is a collection of sensitivity curves for the nonlinear time domain model where the parameter is pressure 
at the eardrum in dB SPL (sound pressure level).}
\label{fig:sens}
\end{figure}

In an isocontour plot, a probe is placed at a specific location on the BM where the time response is measured and analyzed for
input tones covering a range of frequencies.  Figure \ref{fig:sens} shows isointensity curves for $\mathrm{CF} = 10\ \mathrm{kHz}$, 
which corresponds to $\mathrm{CP} = 0.77\ \mathrm{cm}$.  The characteristic place (CP) for a frequency is defined as the location on
the BM of maximal response from a pure tone of that frequency in the fully linear active model ($\gamma = 1$).  The characteristic 
frequency (CF) at a BM location is the inverse of this map.  The left plot is the linear steady state active case.  The parameter 
is the active gain $\gamma$, and for each value of the active gain we get a curve that is a function of the input frequency.  
The value of this function is the ratio $|u|(\mathrm{CP})/P_{\mathrm{e}}$, where $|u|(\mathrm{CP})$ is BM displacement at the 
characteristic place and $P_{\mathrm{e}}$ is pressure at the eardrum.  This is known as sensitivity.  It is basically an output/input 
ratio and gives the transfer characteristics of the ear at that particular active level.  Notice that when $\gamma = 1$, the BM 
at the characteristic place is most sensitive at the corresponding characteristic frequency, but at lower values of the gain, the 
sensitivity peak shifts to lower frequencies.

Analogously, the second plot in Figure \ref{fig:sens} shows isointensity curves for the nonlinear time domain model where now the 
parameter is the intensity of the input stimulus in dB SPL (sound pressure level).  For the time domain, we measure the root-mean-square BM amplitude from
5 ms (to remove transients) up to a certain time $T$.  Note that for high-intensity tones, the model becomes passive while 
low-intensity tones give a more active model.  This shows {\it compression}.  Again, there is a frequency shift of the sensitivity 
peak (about one-half octave) from low to high-intensity stimuli in agreement with \cite{ruggero}, so called half-octave shift.  The 
plot agrees well with Figure 5 in \cite{ruggero}.

\subsection{Complex Stimuli}

\begin{figure}[tb]
\centerline{\epsfig{file=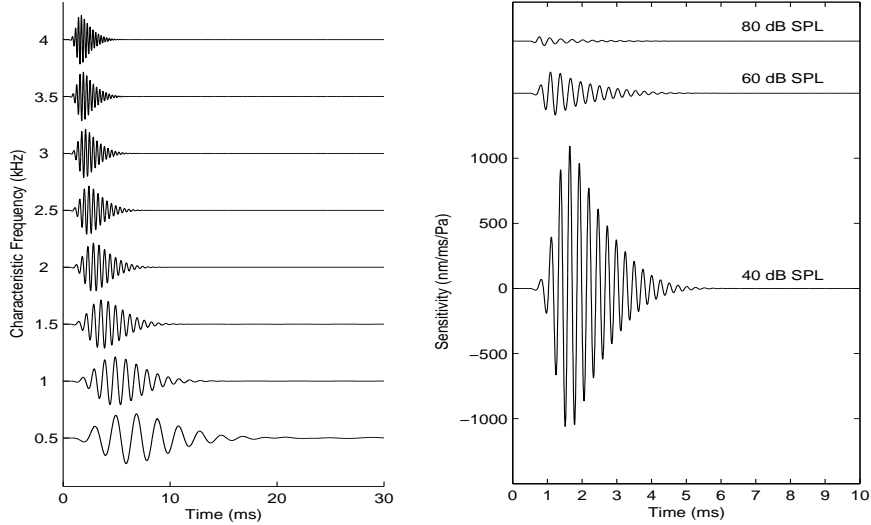,width=330pt,height=200pt}}
\caption{An impulse, or click, lasting 0.1 ms starting at 0.4 ms is input into the nonlinear nonlocal ear model.  The left plot is the BM displacement
time series for various CF's ranging from 0.5-4 kHz.  The right plot is a sensitivity plot for various stimulus intensities
at CF = 6.4 kHz.}
\label{fig:click}
\end{figure}

The first non-sinusoidal input we look at is a click.  In the experiment in the left plot of Figure \ref{fig:click}, 
we put probes at varying characteristic places associated with frequencies ranging from 0.5-4 kHz to measure the time 
series BM displacement.  The click was 40 dB with duration 0.1 ms starting at 0.4 ms.  All responses were normalized to 
amplitude 1.  The plot is similar to Figure 4 in \cite{dengk}.  In the right plot of Figure \ref{fig:click}, a probe
was placed at CP for 6.4 kHz and the time series BM volume velocity was recorded for various intensities and the
sensitivity plotted.  This shows, similar to Figure \ref{fig:sens}, the compression effects at higher intensities.  See
Figure 9 in \cite{ruggero} for a similar plot.

\begin{figure}[tb]
\centerline{\epsfig{file=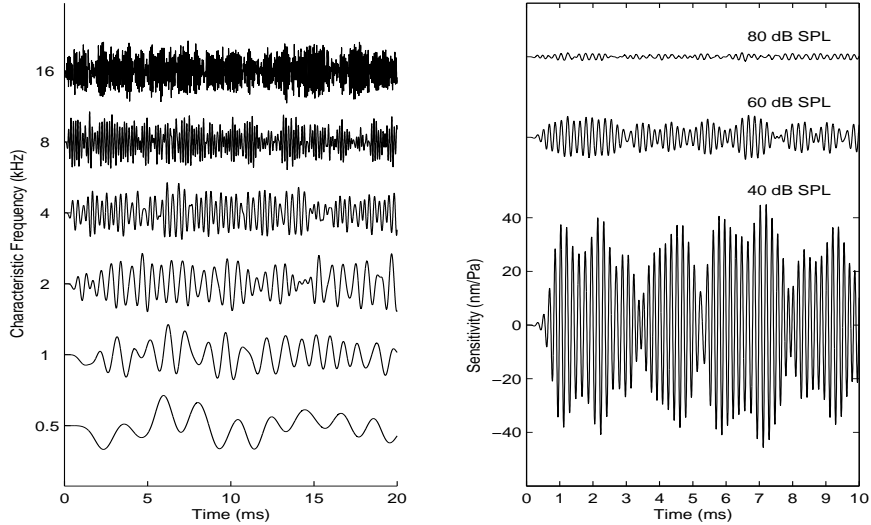,width=330pt,height=200pt}}
\caption{Gaussian noise is input into the ear.  The left plot is the BM displacement times series for various CF's ranging
from 0.5-16 kHz.  The right plot is a sensitivity plot for CF = 6.4 kHz.}
\label{fig:noise}
\end{figure}

The second non-sinusoidal input we explore is Gaussian white noise.  Figure \ref{fig:noise} is similar in all regards
to Figure \ref{fig:click}.  Notice again in the right plot the compression effect.

\bigskip

\subsection{Difference Tones}

\begin{figure}[tb]
\centerline{\epsfig{file=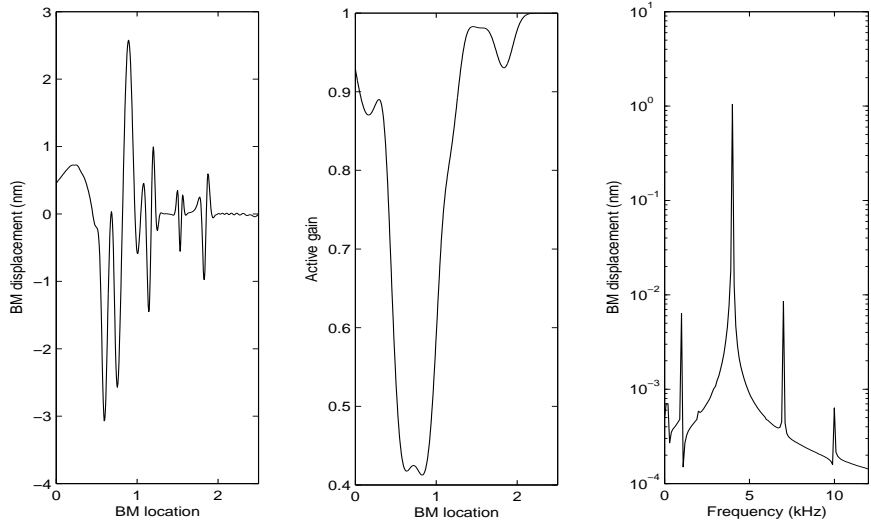,width=330pt,height=200pt}}
\caption{Two sinusoidal tones, 7 and 10 kHz at 80 dB each, are the input.  The left and middle plots are snapshots at 15 ms of BM 
displacement and active gain, respectively.  The right plot is a spectrum plot of the BM displacement time series at CP for 4 kHz.}
\label{fig:cdt}
\end{figure}

Any nonlinear system with multiple sinusoidal inputs will create difference tones.  If 
two frequencies $f_{1}$ and $f_{2}$ are put into the ear, $nf_{1} \pm mf_{2}$ will be created at varying intensities,
where $n$ and $m$ are nonnegative integers.  The cubic difference tone, denoted $f = 2f_{1}-f_{2}$, where 
$f_{1} < f_{2}$, is the most prominent.  Figure \ref{fig:cdt} contains three plots of one experiment.  The experiment 
consists of two sinusoidal tones, 7 and 10 kHz at 80 dB each.  The cubic difference tone is 4 kHz.  The plot on the 
left is the BM profile for the experiment at 15 ms.  We see combination tone peaks at 1.21 cm (CP for 4 kHz), 
1.54 cm (CP for 2 kHz) and 1.85 cm (CP for 1 kHz).  The middle plot shows the snapshot at 15 ms of the active gain parameter,
showing the difference tones getting an active boost.  Finally, the right plot is a spectrum plot of the time series for BM
displacement at 1.21 cm, the characteristic place for 4 kHz.  The cubic difference tone is above 1 nm and can therefore
be heard.

\subsection{Multi-tone Suppression}

\begin{figure}[tb]
\centerline{\epsfig{file=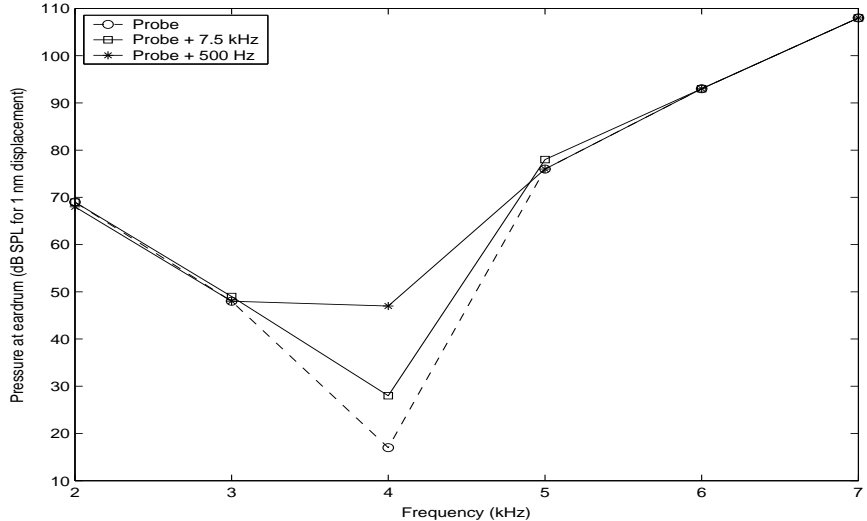,width=330pt,height=200pt}}
\caption{Isodisplacement curves at CP for 4 kHz showing responses to single tones (dashed line w/ circle) and responses to 
the same tones in the presence of high-side and low-side suppressors presented at 80 dB SPL.}
\label{fig:supp}
\end{figure}

\begin{figure}[tb]
\epsfig{file=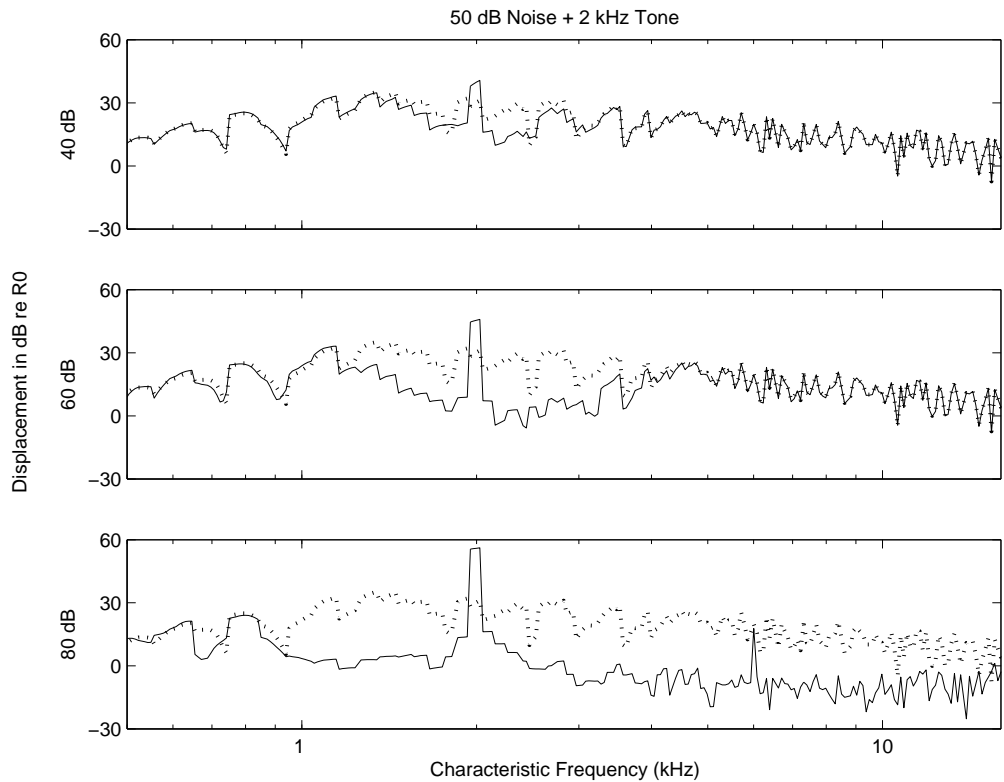,width=380pt}
\caption{Spectrum plots of BM responses for characteristic frequencies along the BM, from 500 Hz to 16 kHz, with 50 dB noise 
and a 2 kHz tone ranging from 40-80 dB.  R0 is the average of the BM response spectrum of 0 dB noise from 0.5-16 kHz.  The solid
line represents noise with tone, the dotted line noise without tone.}
\label{fig:multisupp}
\end{figure}

\begin{figure}[tb]
\epsfig{file=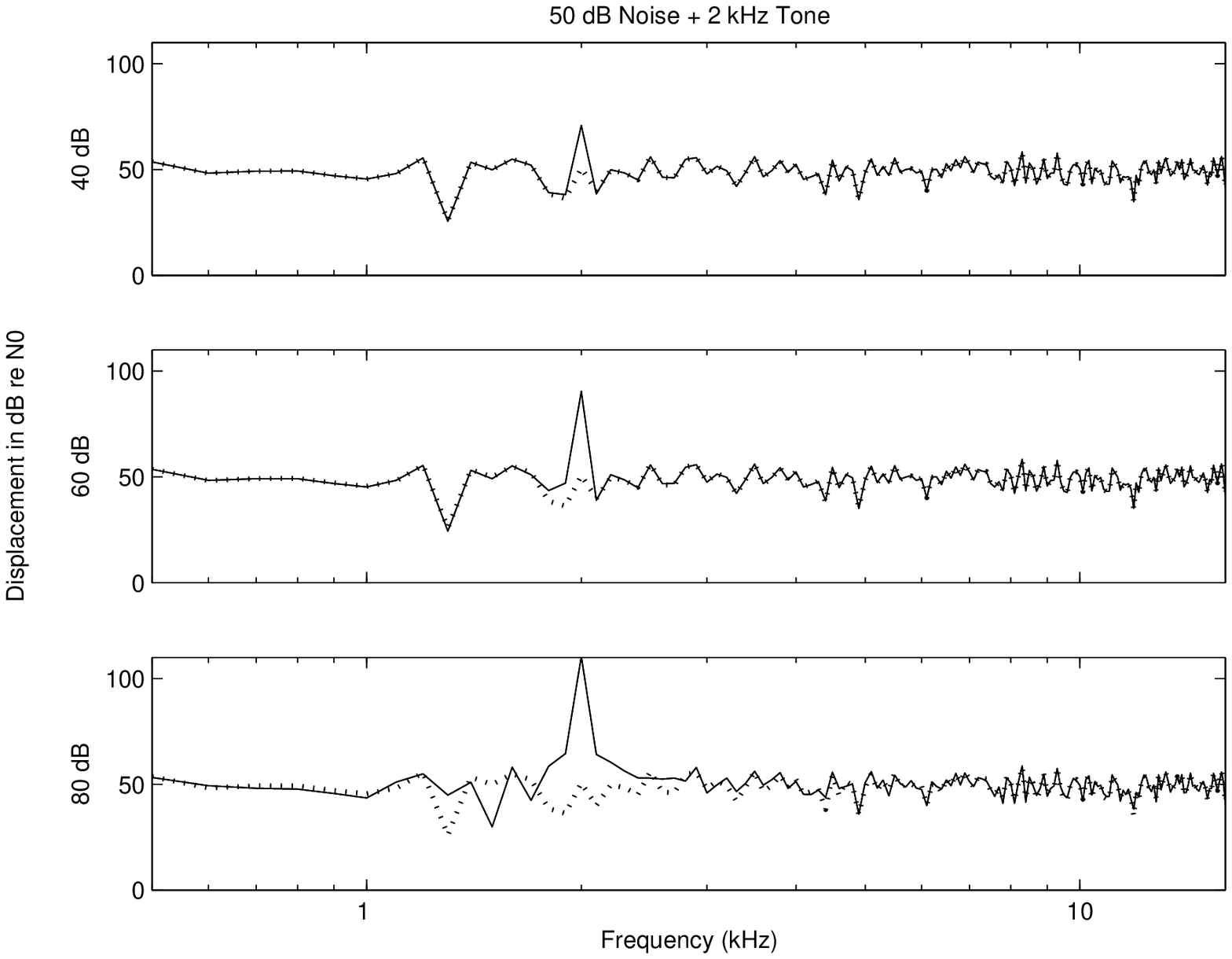,width=380pt}
\caption{Spectrum plots of input signals consisting of 50 dB noise and a 2 kHz tone ranging from 40-80 dB.  N0 is the average
of the spectrum of 0 dB noise from 0.5-16 kHz.  The solid line represents noise with tone, the dotted line noise without tone.}
\label{fig:multisuppfft}
\end{figure}

Two-tone (and multi-tone) suppression is characteristic of a compressive nonlinearity and has been recognized in the
ear \cite{ruggero,dengg,geisler}.  Figure \ref{fig:supp} illustrates two-tone suppression and is a collection of isodisplacement
curves that show decreased tuning in the presence of suppressors and is similar to Figure 16 in \cite{ruggero}.  We placed 
a probe at the CP for 4 kHz (1.21 cm) and input sinusoids of various frequencies.  At each frequency, we record the pressure at 
the eardrum that gives a 1 nm displacement for 4 kHz in the FFT spectrum of the time series response at CP.  The curve
without suppressors is dashed with circles.  We then input each frequency again, but this time in the presence of a low side
(0.5 kHz) tone and high side (7.5 kHz) tone, both at 80 dB.  Notice the reduced tuning at the CF.  Also notice the asymmetry of 
suppression, which shows low side is more suppressive than high side, in agreement with \cite{geisler}.

For multi-tone suppression, we look at tonal suppression of noise.  In Figure \ref{fig:multisupp}, 
for each plot, a probe was placed at every grid point along the BM and the time response was measured from 15 ms up to 25 ms.  
The signal in each consisted of noise at 50 dB with a 2 kHz tone ranging from 40 dB to 80 dB (top to bottom).
An FFT was performed for each response and its characteristic frequency amplitude was recorded and plotted in decibels 
relative to the average of the response spectrum of 0 dB noise from 0.5-16 kHz.  We see suppression of all frequencies, with again low-side suppression stronger 
than high-side suppression.  Figure \ref{fig:multisupp} is qualitatively similar to Figure 3 in \cite{dengg}.  It is useful to 
compare this figure with Figure \ref{fig:multisuppfft}.  This figure is the same as Figure \ref{fig:multisupp}, except we
do an FFT of the input signal at the eardrum.  Comparing these two figures shows that we have a new spectral transform that can
be used in place of an FFT in certain applications, for example 
signal recognition and noise suppression.

\section{Conclusions}

We studied a two-dimensional nonlinear nonlocal variation of the 
linear active model in \cite{neelykim}.
We then developed an efficient and accurate numerical method and used this method to explore nonlinear effects
of multi-tone sinusoidal inputs, as well as clicks and noise.  We showed numerical results illustrating compression,
multi-tone suppression and difference tones.  
The model reached agreement with experiments \cite{ruggero} and  
produced a novel nonlinear spectrum.  In future work, we will analyze the model 
responses to speech and resulting spectra for speech 
recognition.   
It is also interesting to study 
the inverse problem \cite{sondhi}
of finding efficient and automated ways 
to tune the model to different physiological data. 
Applying the model to 
psychoacoustic signal processing \cite{jxinPDE} will be another 
fruitful line of inquiry.
\vspace{.1 in} 

\begin{ack}
The work was partially supported by NSF grant ITR-0219004.
J. X. would like to acknowledge a fellowship from the John Simon Guggenheim Memorial Foundation, and a
Faculty Research Assignment Award at UT Austin.
\end{ack}

\begin{appendix}

\section{Appendix: Convergence of Iterative Scheme (\ref{eq:iteration})}

We need the following Lemma:
\begin{lem}
If
\[M = \left[ \begin{array}{cc}
                A & -A \\
		B & -B
	     \end{array}
      \right]
\]
then every non-zero eigenvalue of $M$ is an eigenvalue of $A-B$.
\end{lem}
\begin{pf}

Let $\lambda$ be a non-zero eigenvalue of $M$ with non-trivial eigenvector $\vec{x} = (\vec{x}_{1},\vec{x}_{2})$.  Thus,
$M\vec{x} = \lambda\vec{x}$ gives
\begin{equation}
   A(\vec{x}_{1}-\vec{x}_{2}) = \lambda\vec{x}_{1} \label{equ:A1}
\end{equation}
\begin{equation}
   B(\vec{x}_{1}-\vec{x}_{2}) = \lambda\vec{x}_{2} \label{equ:A2}
\end{equation}
Subtracting the two equations, we have
\[(A-B)(\vec{x}_{1}-\vec{x}_{2}) = \lambda(\vec{x}_{1}-\vec{x}_{2})\]
Now, if $\vec{x}_{1}-\vec{x}_{2} = 0$, then from \ref{equ:A1} and \ref{equ:A2} above and $\lambda \neq 0$, we have 
$\vec{x}_{1} = \vec{x}_{2} = 0$.  But this means $\vec{x} = 0$, which is a contradiction.  Thus, $\lambda$ is an 
eigenvalue of $A-B$ with non-trivial eigenvector $\vec{x}_{1}-\vec{x}_{2}.$\qed
\end{pf}

\begin{thm}
There exists a constant $C > 0$ such that if $\Delta{t} < C$, then
\[\rho(\Lp^{-1}\La) < 1\]
where $\rho$ is the spectral radius.  Thus, the iterative scheme converges.
\end{thm}
\begin{pf}

By the above lemma applied to (\ref{eq:lpinvla}), with constant $\gamma$, we have
\begin{eqnarray*}
  \sigma(\Lp^{-1}\La) & \subset & \gamma\sigma(D^{-1}P_{4}-\tilde{M}_{2}^{-1}P_{3}D^{-1}P_{4}) \\
                      & =       & \gamma\sigma[(I - \tilde{M}_{2}^{-1}P_{3})D^{-1}P_{4}]
\end{eqnarray*}
where $\sigma$ denotes spectrum.  Thus, we have
\begin{eqnarray*}
  \rho(\Lp^{-1}\La) & \leq & \gamma||(I-\tilde{M}_{2}^{-1}P_{3})W^{-1}\Ds^{-1}P_{4}||_{2} \\
                    & \leq & \gamma||(I-\tilde{M}_{2}^{-1}P_{3})W^{-1}||_{2}||\Ds^{-1}||_{2}||P_{4}||_{2}
\end{eqnarray*}
Now, let $(\lambda,\vec{x})$ be the eigen-pair of $\Ds$ with $\lambda$ the smallest eigenvalue and $||\vec{x}|| = 1$.
Note that $\lambda > 0$ since $\Ds$ is positive definite.  Thus, we have $1/\lambda$ is the largest eigenvalue of $\Ds^{-1}$,
which gives
\[||\Ds^{-1}||_{2} \leq 1/\lambda\]
Thus, using the definition of $\Ds$ from (\ref{eq:dsdef}), we have
\begin{eqnarray*}
  \lambda &    = & \vec{x}^{T}\Ds\vec{x}\\
          &    = & \vec{x}^{T}\{2\alpha \Mfs + [2M_{1}+P_{1}+P_{3}(I-\tilde{M}_{2}^{-1}P_{3})]W^{-1}\}\vec{x}\\
	  & \geq & \vec{x}^{T}\{[2M_{1}+P_{1}+P_{3}(I-\tilde{M}_{2}^{-1}P_{3})]W^{-1}\}\vec{x}\\
	  & \geq & \min\{[2m_{1}+p_{1}+p_{3}(1-\tilde{m}_{2}^{-1}p_{3})]w^{-1}\}
\end{eqnarray*}
where lowercase represents diagonal entries.  The third line above follows from $2\alpha \Mfs$ being positive definite.
Finally, we have
\begin{eqnarray*}
  \rho(\Lp^{-1}\La) & \leq & \gamma||(I-\tilde{M}_{2}^{-1}P_{3})W^{-1}||_{2}||\Ds^{-1}||_{2}||P_{4}||_{2} \\
                    & \leq & \gamma\frac{\max[(1-\tilde{m}_{2}^{-1}p_{3})w^{-1}]\max(p_{4})}{\min\{[2m_{1}+p_{1}+p_{3}(1-\tilde{m}_{2}^{-1}p_{3})]w^{-1}\}}
\end{eqnarray*}
For $\Delta{t}$ small enough, we have convergence.\qed
\end{pf}

With our parameters, for convergence it is {\it sufficient} that $\Delta{t} \leq 0.0008$.  In practice, however, convergence 
is seen for $\Delta{t}$ as large as $0.01$.
\end{appendix}

\end{document}